%% file: main.tex
\documentclass[sigconf]{acmart}
\pdfoutput=1

\DeclareMathOperator*{\argmax}{arg\,max}
\usepackage{subcaption}
\usepackage{url}
\usepackage{graphicx}

\AtBeginDocument{%
  }

\copyrightyear{2024}

\setcopyright{none}
\renewcommand\footnotetextcopyrightpermission[1]{}

\acmConference[]{}{}{}

\begin{document}
\pagestyle{empty}

\title{SwipeGANSpace: Swipe-to-Compare Image Generation via Efficient Latent Space Exploration}

\author{Yuto Nakashima}
\email{yuto-nakashima@g.ecc.u-tokyo.ac.jp}
\affiliation{%
  \institution{The University of Tokyo}
  \country{Japan}
  \city{Tokyo}
}

\author{Mingzhe Yang}
\email{mingzhe-yang@g.ecc.u-tokyo.ac.jp}
\affiliation{%
  \institution{The University of Tokyo}
  \country{Japan}
  \city{Tokyo}
}

\author{Yukino Baba}
\email{yukino-baba@g.ecc.u-tokyo.ac.jp}
\affiliation{%
  \institution{The University of Tokyo}
  \country{Japan}
  \city{Tokyo}
}

\renewcommand{\shortauthors}{Nakashima et al.}

\begin{abstract}
 Generating preferred images using generative adversarial networks (GANs) is challenging owing to the high-dimensional nature of latent space. In this study, we propose a novel approach that uses simple user-swipe interactions to generate preferred images for users. To effectively explore the latent space with only swipe interactions, we apply principal component analysis to the latent space of the StyleGAN, creating meaningful subspaces. We use a multi-armed bandit algorithm to decide the dimensions to explore, focusing on the preferences of the user. Experiments show that our method is more efficient in generating preferred images than the baseline methods. Furthermore, changes in preferred images during image generation or the display of entirely different image styles were observed to provide new inspirations, subsequently altering user preferences. This highlights the dynamic nature of user preferences, which our proposed approach recognizes and enhances.
\end{abstract}




\maketitle

\section{Introduction}
\input{source/body/Introduction}

\section{Related Work}
\input{source/body/RelatedWork}

\section{SWIPE-TO-COMPARE IMAGE GENERATION}
\input{source/body/SwipeToCompareImageGeneration}

\section{Simulation experiments}
\input{source/body/SimulationExperiment}

\section{User experiments}
\input{source/body/UserExperiment}

\section{Discussion}
\input{source/body/Discussion}

\section{Design implications}
\input{source/body/DesignImplication}

\section{Conclusion}
\input{source/body/Conclusion}

\begin{acks}
  This research is part of the results of Value Exchange Engineering, a joint research project between R4D, Mercari Inc., and the RIISE.
\end{acks}

\bibliographystyle{ACM-Reference-Format}
\bibliography{reference}

\appendix

\end{document}

%% file: source/body/Introduction.tex
\begin{figure*}[tb]
  \centering
  \includegraphics[width=.8\linewidth]{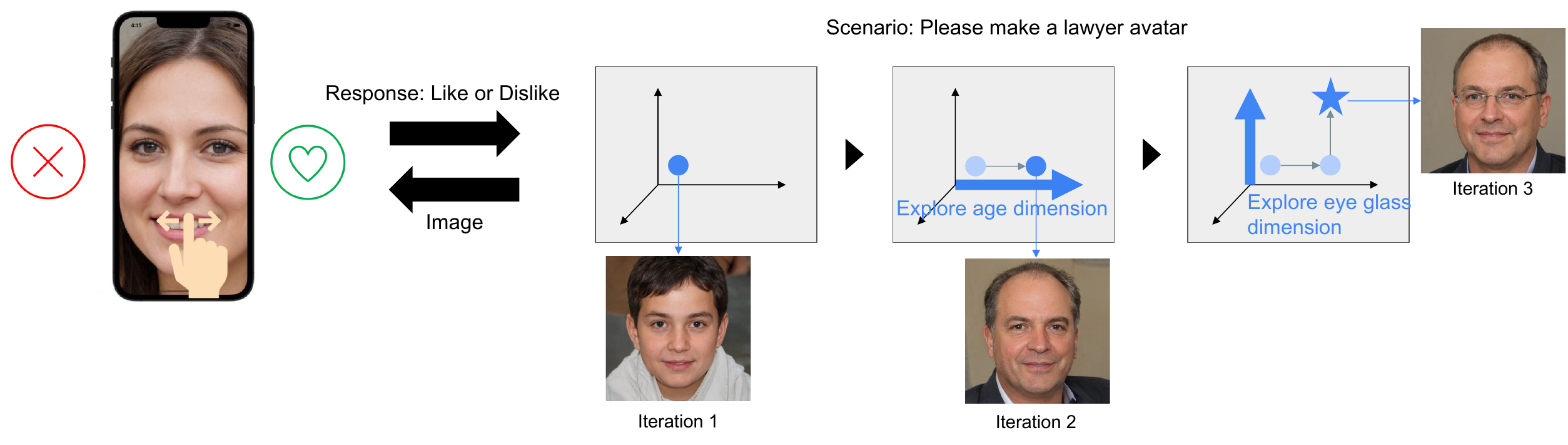}
 \caption{Overview of image generation through swipes. This example illustrates how our system generates a lawyer avatar based on user swipes. By interpreting continuous swipe feedback, the system dynamically adjusts the displayed image to match user preferences. Internally, it evaluates dimensions such as age or the presence of glasses. For instance, in Iteration $2$, the system updates the age of the avatar based on estimated user preferences. By the third iteration, in response to feedback, glasses are added to the avatar.}

  \label{fig: teaser-figure}
\end{figure*}

Generative adversarial networks (GANs) facilitate the generation of high-quality images. One of the notable features of GANs is their ability to allow users to modify these images by adjusting latent variables~\cite{GANSpace}. It is essential to select the correct latent variables to obtain the desired results; however, choosing the right latent variables is challenging given the high dimensionality of the latent space of GANs. Previous research suggested using multiple sliders~\cite{multiple_slider, InterfaceGAN, GANSpace, SeFa} or image editing tools~\cite{Generativevisualmanipulationonthenaturalimagemanifold} to allow users to modify the images. These approaches are sometimes inconvenient for users to apply, especially on smartphones, where the limited screen space complicates the use of sliders and editing tools. Although text-to-image techniques offer a potential solution to generate preferred images, users often find it challenging to design appropriate prompts.

In this study, we propose a method that allows users to generate images based on their preferences with minimal interaction. Our system uses \textit{swipe} actions, commonly used in smartphone interfaces. An overview of our approach is illustrated in Figure~\ref{fig: teaser-figure}. Given the familiarity many users have with swipe-based applications such as Tinder\footnote{https://tinder.com\label{tinder}}, we design our user interface to present one image at a time, relying on swipe gestures for feedback. Instead of extensive user input, our system estimates preferences based on these swipes. When the system displays an image, users swipe to indicate their preference. Based on this feedback, the system generates new images that better match these preferences. With repeated interactions, the system becomes more adept at producing images aligned with what the user likes and dislikes.

To generate preferred images using swipe interactions, we need methods that enhance the efficiency of exploring latent variables and operate on more limited feedback compared with previous approaches. The advent of StyleGAN~\cite{StyleGAN} has enabled the exploitation of meaningful latent spaces, such as style information. Additionally, GANSpace~\cite{GANSpace} demonstrated that when principal component analysis (PCA) is applied to the latent space of StyleGAN, it becomes feasible to identify principal components that significantly influence image appearance. Inspired by these findings, we apply PCA to StyleGAN latent space, establishing a subspace built on principal components that markedly modify image appearance. Within this subspace, we use Bayesian optimization to enhance the efficiency of latent variable exploration. 

It is essential to maintain a large subspace to preserve the expressive capability of GAN. However, increasing the dimensionality of the subspace has its attendant challenges: Bayesian optimization becomes less effective in high-dimensional spaces, making it challenging to solely rely on Bayesian optimization to generate preferred images. To address this problem, we employ a multi-armed bandit algorithm. This algorithm dynamically identifies the dimensions that should be explored, focusing on those particularly relevant to the user, thereby enhancing the efficiency of the search process.

To evaluate the efficiency of our proposed method in generating user-preferred images, we performed both simulation and user experiments. The simulation results indicate that, in cases of large subspace dimensionality, our method is more efficient than the baseline method. 
The result of user experiments also show that our  method can generate preferred images more efficient than the baseline method. Moreover, we observed that user preferences can shift when comparing images. Our proposed method effectively accommodated these changes, promoting a broader scope for user exploration.

Our contributions are summarized as follows:
\begin{itemize}
\item We introduce a novel user interface for generating user-preferred images that are simply controlled with a swipe. Users can easily generate these images by repeatedly swiping, similar to how they would use a matchmaking app.

\item To efficiently explore the GAN latent space using minimal feedback from swipe interactions, we devise an approach that integrates Bayesian optimization with a multi-armed bandit algorithm. This algorithm dynamically determines the dimensions within the subspace that are of interest to the user.

\item Through the user study, we confirmed the efficiency of our proposed method in generating preferred images. Furthermore, we observed a gradual shift in user preferences when presented with pairwise comparisons. Our method not only accommodates shifts in user preferences but also allows users to reconsider and adjust their choices. 

\end{itemize}

%% file: source/body/RelatedWork.tex
\subsection{Generating preference images from GANs}
Recently, there has been a surge in interest in image-generation models such as variational autoencoders~\cite{VAE}, GANs~\cite{GAN, StyleGAN, CycleGAN, StackGAN}, autoregressive models~\cite{pixelRNN, PixelCNN, DRAW, parti}, and diffusion models~\cite{diffusion_model, stable_diffusion,imagen, dall-e2}. In particular, GANs can produce high-quality images, which has led to an increased focus on image-generation methods  tailored for user preferences. One approach to generating preference images is to use conditional GANs~\cite{pix2pix, CycleGAN, Interactiveexample-basedterrainauthoringwithconditionalgenerativeadversarialnetworks, Real-timeuser-guidedimagecolorization, Faceshop:Deepsketch-based}. However, these approaches required custom network architectures and training data to control particular applications. Unlike previous studies, our approach uses a pre-trained model and does not require further training.

Research on image-generation control has been conducted by analyzing the latent space of pre-trained deep generative models. By controlling the latent space, preference images can be generated by drawing or coloring on a blank canvas~\cite{Generativevisualmanipulationonthenaturalimagemanifold, Semanticphotomanipulationwithagenerativeimageprior}, adjusting multiple sliders~\cite{InterfaceGAN, GANSpace, SeFa, UnsupervisedDiscoveryofInterpretableDirectionsintheGANLatentSpace}. In contrast to previous studies, our approach focuses on simplifying user input by introducing \textit{swiping}, the simplest method of interaction. Previous studies have controlled attributes in the latent space by providing users with canvas, sliders, and so on, and required them to draw sketches or adjust multiple sliders, which can be difficult for users. In contrast, our method can generate preferred images using only swiping actions. Furthermore, our method is especially beneficial in cases where users do not have or are unable to express their preferences clearly. Existing studies assume that users have a clear idea of what they want. However, our method does not make this assumption, allowing them to generate any image they prefer even when they cannot express their intentions accurately. Without a clear object, a preferred image can be created by simply selecting the preferred image.

\subsection{Human-in-the-loop optimization}
To generate preferred images using only swiping actions, it is necessary to search effectively in the latent space. Therefore, we used human-in-the-loop optimization, which is an efficient way to explore parameters when editing images. Human-in-the-loop optimization involves humans as computational resources in iterative optimization computations. It is advantageous to include humans in iterations, especially when optimizing for human preferences. Owing to these advantages, it has been widely used in recent years~\cite{TakingtheHumanOutoftheLoop,DesigningEngagingGamesUsingBayesianOptimization, multiple_slider, CrowdsourcingInterfaceFeatureDesign, Generative-melody-composition, sequential_line, sequential_garally}. Bayesian optimization for human-in-the-loop is a collaborative problem-solving approach between humans and computers. The computer samples potential solutions to optimize the objective function and requests humans to evaluate those candidates. This process is repeated until a desirable solution is found. In this approach, it is particularly important to ask questions that humans can easily answer and to reduce the number of queries.

It is difficult to obtain consistent absolute evaluations for a single design. Therefore, using relative evaluations on multiple candidates is recommended~\cite{HowtoAnalyzePairedComparisonData, BayesianInteractiveOptimizationApproachtoProceduralAnimationDesign}. Brochu et al.~\cite{ActivePreferenceLearning} proposed preferential Bayesian optimization~\cite{PBO}, which uses Bayesian optimization with preference comparison data, and various extensions have been made since then~\cite{multiple_slider, sequential_garally, sequential_line}. For typical parameter spaces (around $10$ dimensions), interfaces such as slider adjustments~\cite{sequential_line} and N-pair comparisons~\cite{sequential_garally} have been proposed. However, these studies have shown that these sequential line search methods are not effective because the latent space of GANs is high-dimensional ($512$-dimensions). Moreover, they require users to respond to more complicated queries such as multiple slider adjustments and image editing, which are cognitively burdensome. Our study proposes generating preference images using only swipe operations, even in a high-dimensional search space. We present two steps to generate images using only swipe operations. First, we apply PCA to StyleGAN latent space and created a subspace composed of directions that significantly alter the appearance of images, thus reducing the exploration range. Second, we use a multi-armed bandit algorithm to determine the exploration dimension, allowing focused exploration in dimensions of interest to the user.

%% file: source/body/SwipeToCompareImageGeneration.tex
\begin{figure*}[tb]
  \centering
  \includegraphics[width=.8\linewidth]{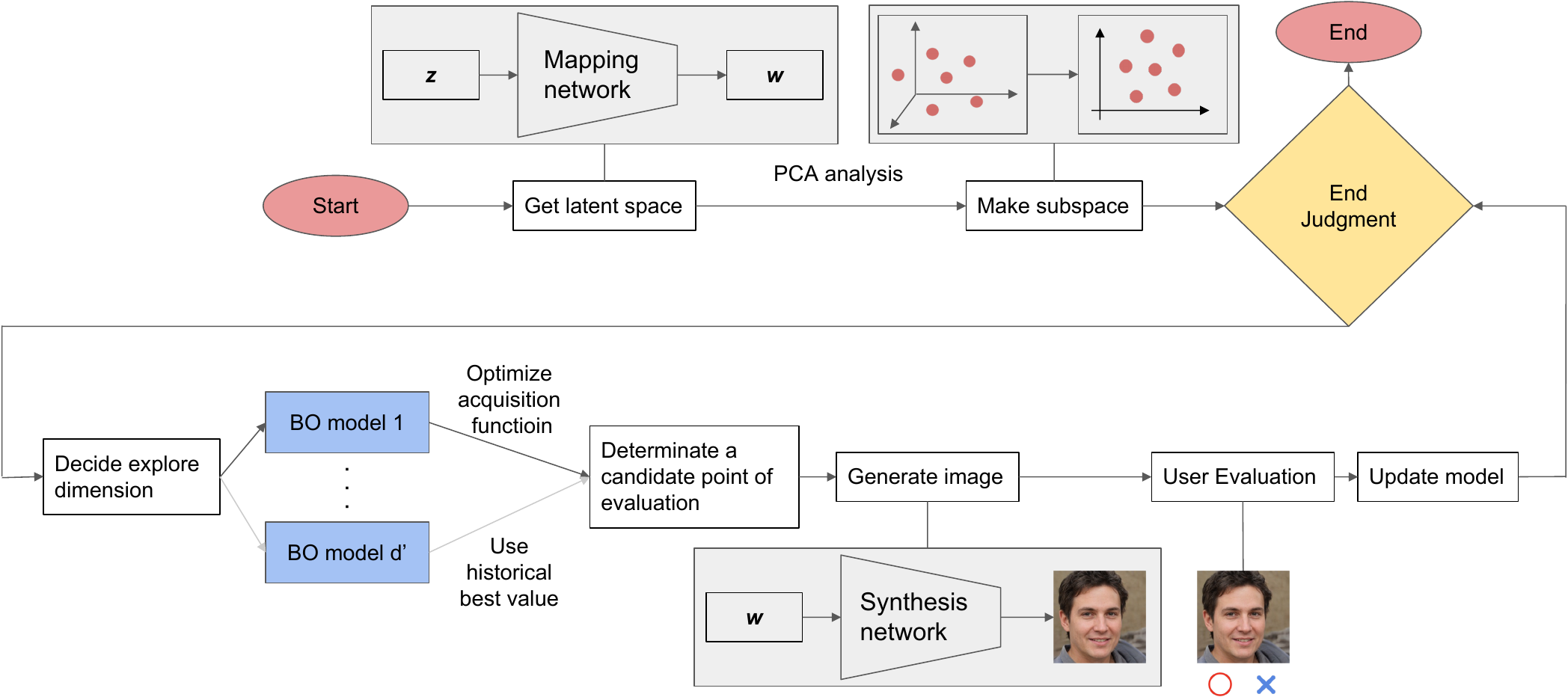}
  \caption{Overview of the proposed method: We initially employ PCA to derive a subspace from the latent space of StyleGAN. Next, the system identifies a key dimension using a multi-armed bandit algorithm and performs Bayesian optimization within the latent space of this dimension. The resulting latent variables are transformed into an image and presented to the user. Based on user feedback, the model is updated accordingly.}
  \label{fig: proposed method overview}
\end{figure*}

\subsection{Swipe interactions for image generation}
We introduce a new user interface, illustrated in Figure~\ref{fig: teaser-figure}, designed to generate user-preferred images using simple swipe actions. This interface captures user preferences by continuously processing swipe interactions to generate the preferred images. Inspired by smartphone matchmaking applications such as Tinder, our interface displays a single image at a time. Users are asked to perform pairwise comparisons by comparing the currently displayed image with the previous one: swiping right indicates a preference for the current image, whereas swiping left favors the earlier one. Depending on this feedback, subsequent images are generated and presented. This process continues until the image aligns with the preference of the user.

\subsection{Proposed method for efficient latent-space search}
\subsubsection{Image-generation model.}
We use StyleGAN~\cite{StyleGAN} to generate high-quality images and to support latent-space encoding style information. We sample a random vector $\boldsymbol{z}$ from the latent space of StyleGAN and use it as the input to the mapping network to obtain the intermediate latent variable $\boldsymbol{w}$, which contains the information about the style. The synthesis network then takes $\boldsymbol{w}$ as the input to generate images, allowing control over the style of the generated images.

\subsubsection{Objective function.}
To generate preference images, it is necessary to search for optimal latent variables. We consider an optimization problem that takes human preferences as an objective function and intermediate latent variables $\boldsymbol{w} \in W$ in StyleGAN as search variables. Let the synthesis network of StyleGAN be $s\colon W \to X$, where $W \subset \mathbb R^{d}$ is a high-dimensional space such as $d=512$, and $X \subset \mathbb{R}^{h \times w}$ is an image data space such that $h=w=1{,}024$. A user attempts to generate image data $\boldsymbol{x} \in X$ using this generation model, $s$, but only the vector $\boldsymbol{w} \in W$ can be controlled. The preference for image data $\boldsymbol{x}$ is determined by the function $g\colon X \to \mathbb R$, which is unknown to the system. The goal is to solve the following optimization problem to generate images with high preferences:
\begin{equation}
\label{equation: objecive equation}
   \boldsymbol w^{*} = \arg \max_{\boldsymbol w \in W} g\left(s\left(\boldsymbol w \right)\right)
\end{equation}
where $g\left(s\left(\boldsymbol w\right)\right)$ can be computed for any point $\boldsymbol w$, but $g\left(\cdot\right)$ is expensive to evaluate, and it is important to reduce the number of function evaluations as much as possible. 

\subsubsection{Bayesian optimization with pairwise comparison data.}
We use Bayesian optimization~\cite{TakingtheHumanOutoftheLoop} to evaluate the objective function and solve Equation~\eqref{equation: objecive equation}. We consider Bayesian optimization by defining a composite function $f= g \circ s$ of $g$ and $s$, where $f$ is the objective function. Instead of directly using the value of $f$, our proposed method incorporates the pairwise comparison results provided by the user. By applying the existing method~\cite{PreferenceLearningwithGaussianProcesses}\footnote{We used the implementation of BoTorch's Pairwise GP:\url{https://botorch.org/api/models.html}}, the results of the pairwise comparisons are converted into real values based on the Bayesian estimation to compute $f$. We set the hyperparameters with default values. When querying users, latent variables $\boldsymbol{w}$ are converted to images $\boldsymbol{x} \in X$ using the synthesis network in StyleGAN.

\subsubsection{Dimension reduction.}
One issue with pairwise comparisons is that the number of inquiries to humans increases due to the small amount of user preference information obtained from each iteration. Elena et al.~\cite{PCA_BO} demonstrated that when performing Bayesian optimization in a high-dimensional space, the search can proceed efficiently by applying PCA to the search space, reducing the real search space. In GANSpace~\cite{GANSpace}, PCA is applied to the latent space of StyleGAN to identify principal components that significantly change the appearance of images. Inspired by them, we applied PCA to the latent space of StyleGAN and explored in a subspace composed of principal components only to streamline the search for latent variables.

Figure~\ref{fig: proposed method overview} shows an overview of the proposed method. First, $N$ vectors $\boldsymbol{z} \in \mathbb R^{d}$ are randomly sampled from the standard normal distribution. Next, by inputting the sampled $\boldsymbol{z}$ into the StyleGAN mapping network, the intermediate latent variable $\boldsymbol{w} \in W \subset \mathbb{R}^{d}$ is obtained. Using $\boldsymbol{w}$, PCA is performed to create a $d^{\prime}$-dimensional subspace $W^{\prime} \subset \mathbb{R}^{d^{\prime}}$ with the top $d^{\prime}$ principal components in terms of the contribution rate. Thereafter, Bayesian optimization is performed within subspace $W^{\prime}$. The evaluation point proposed $\boldsymbol {w}^{\prime} \in W^{\prime}$ in Bayesian optimization has a dimension of $d^{\prime} \neq d$, which makes it impossible to directly input it into the StyleGAN synthesis network. To solve this, the inverse transformation of PCA is used to map $\boldsymbol {w}^{\prime}$ onto the latent space $W$ of StyleGAN, which is then inputted into the synthesis network. The generated image is then presented to the user and a pairwise comparison result with the last generated image is obtained.

\subsubsection{Identification of key dimensions.}
The size of the subspace should be kept large to maintain the capacity of GAN for expression. However, if the dimensionality of the subspace is increased, Bayesian optimization will not be effective in high-dimensional spaces~\cite{highdimensionalBOsurvey}, making it difficult to generate preferred images using Bayesian optimization alone. To enhance the search efficiency, we apply a multi-armed bandit algorithm to dynamically determine the dimensions for exploration. Specifically, a Bayesian optimization model is prepared for each dimension in advance, the multi-armed bandit algorithm is used to select the key dimension for the search, and candidate points are determined using the Bayesian optimization model. The candidate points are then converted into images to obtain user feedback. The multi-armed bandit model is updated using the results of image comparisons from users.

We apply the upper confidence bound (UCB) algorithm~\cite{UCB}, a common multi-armed bandit algorithm, due to its notable performance and simplicity in implementation. The UCB algorithm calculates the UCB score for each dimension in each iteration and searches for the dimension with the highest UCB score. At the $t$-th iteration, assuming that the number of times that the $i$-th dimension is searched so far is $N_{i}$ and the estimated value of the reward is $\hat{r}_{i}$, the UCB score for the $i$-th dimension $U_{i}$ is calculated by 
\begin{equation}
    U_{i} = \hat{r}_{i} + \sqrt{\frac{\alpha \log t}{N_{i}}}
\end{equation}
where $\alpha$ is a hyperparameter that adjusts the trade-off between exploration and exploitation.\footnote{In the experiments, we employed $\alpha=0.5$ which had a good balance between exploration and exploitation.}

$U_{i}$ is used to select the largest dimension $i^*$:
\begin{equation}
    i^* = \argmax_i U_{i}.
\end{equation}
The multi-armed bandit model is then updated as follows:
\begin{equation}
    \hat{r}_{i} = \frac{\sum_{j=1}^{N_{i}} r_{i,j}}{N_{i}}
\end{equation}
where $r_{i,j} \in \{0, 1\}$ is the reward when the $i$-th dimension is selected for the $j$-th time. If the last generated image is selected, then $r_{i,j}=0$, and if the current generated image is selected, then $r_{i,j}=1$. For the selected dimension $i^*$, Bayesian optimization is used to determine the $i^*$ components of the candidate points. Let $A$ be the acquisition function and $w_{i^{*}}$ be the $i^*$-th component of the candidate point:
\begin{equation}
w_{i^{*}} = \argmax_{w \in R} A\left(w\right).
\end{equation}
For dimensions other than $i^*$, we assign the best value of the Bayesian optimization model for each dimension based on previous observations. Let $w_0, w_1,\dots , w_{d^\prime}$ be the best values for each dimension and $\boldsymbol{w^{\prime}_t}=[w_0, w_1, \dots , w_{i^{*}}, \dots , w_{d^\prime}]$ be the candidate evaluation point at the $t$-th iteration. The candidate point $\boldsymbol{w^{\prime}_t}$ is transformed into an image using the inverse transform of PCA and the synthesis network to query the user.

%% file: source/body/SimulationExperiment.tex
\subsection{Experimental setup}
\begin{figure}[tb]
  \centering
  \includegraphics[width=\linewidth]{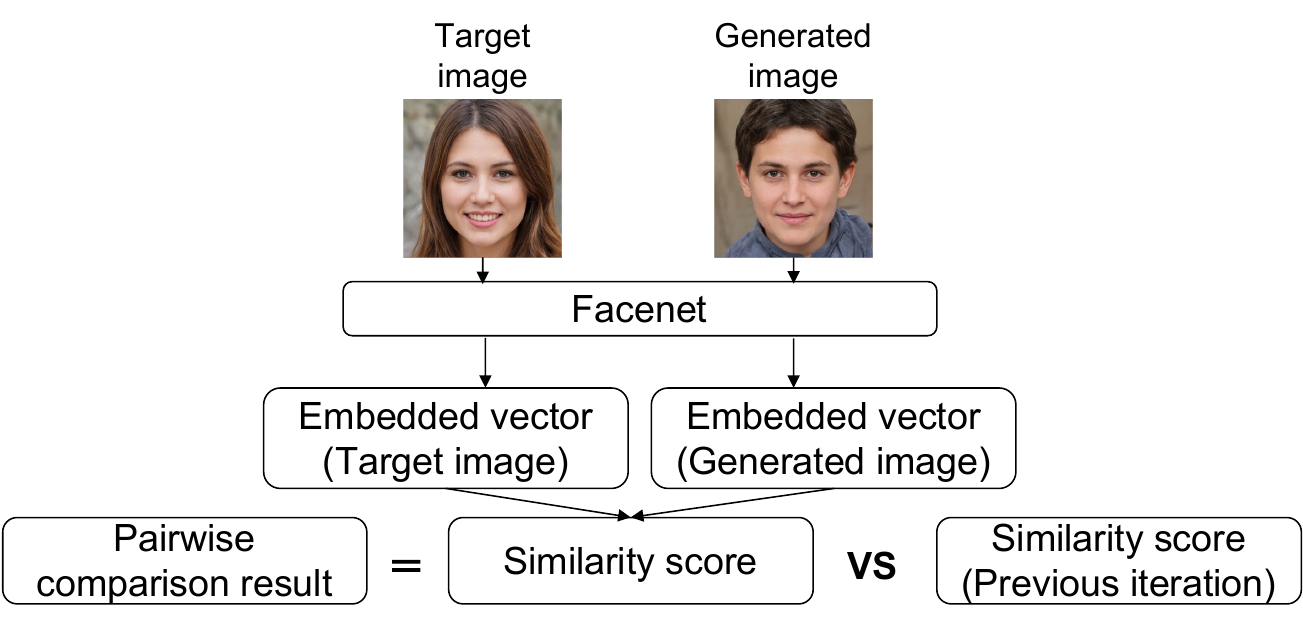}
  \caption{Creating pairwise comparison for simulation experiments. There are two steps to create a pairwise comparison. First, the system obtains the embedding vectors of the target and generates images using Facenet. Second, the pairwise comparison results are created by taking the cosine similarity of these two vectors and comparing this similarity to the similarity in the previous iteration.}
  \label{fig: Overview of our simulation experiment}
\end{figure}

To evaluate whether the proposed method efficiently approaches preferred images assuming that the user provides a stable response, we conducted a simulation experiment. We compared our method (BanditBO) with the baseline method (SimpleBO) without the bandit algorithm from the proposed method. Figure~\ref{fig: Overview of our simulation experiment} shows an overview of the simulation experiment. We simulated a situation in which a user has a specific goal of creating an image using our system. To generate the pairwise comparison results, we predefined a target image, calculated the similarity between the generated and target images, and compared it with the results of previous iterations.

The pairwise comparison results were then input into the Bayesian optimization system. The dimensionality of the subspace $W^{\prime}$ was set to $d^{\prime} \in \{4, 8, 16\}$. Ten target images were randomly selected from the search space for this experiment. We used a pre-trained StyleGAN model that was originally trained on the Flickr-Faces-HQ (FFHQ) dataset~\cite{StyleGAN}. The FFHQ dataset is designed as a GAN benchmark and comprises more than $70{,}000$ images of human faces sized $1{,}024\times1{,}024$, covering various ages, gender, races, and facial expressions.

We used FaceNet~\cite{Facenet} to calculate the similarity between two images. FaceNet is a CNN model for facial recognition that converts facial images into embedded vectors. Using FaceNet, the embedding vectors of the generated and target images are obtained and the cosine similarity between the embedding vectors is calculated.

\subsection{Results}
\begin{figure*}[tb]
  \centering
  \begin{subfigure}{0.32\linewidth}
    \includegraphics[width=\linewidth]{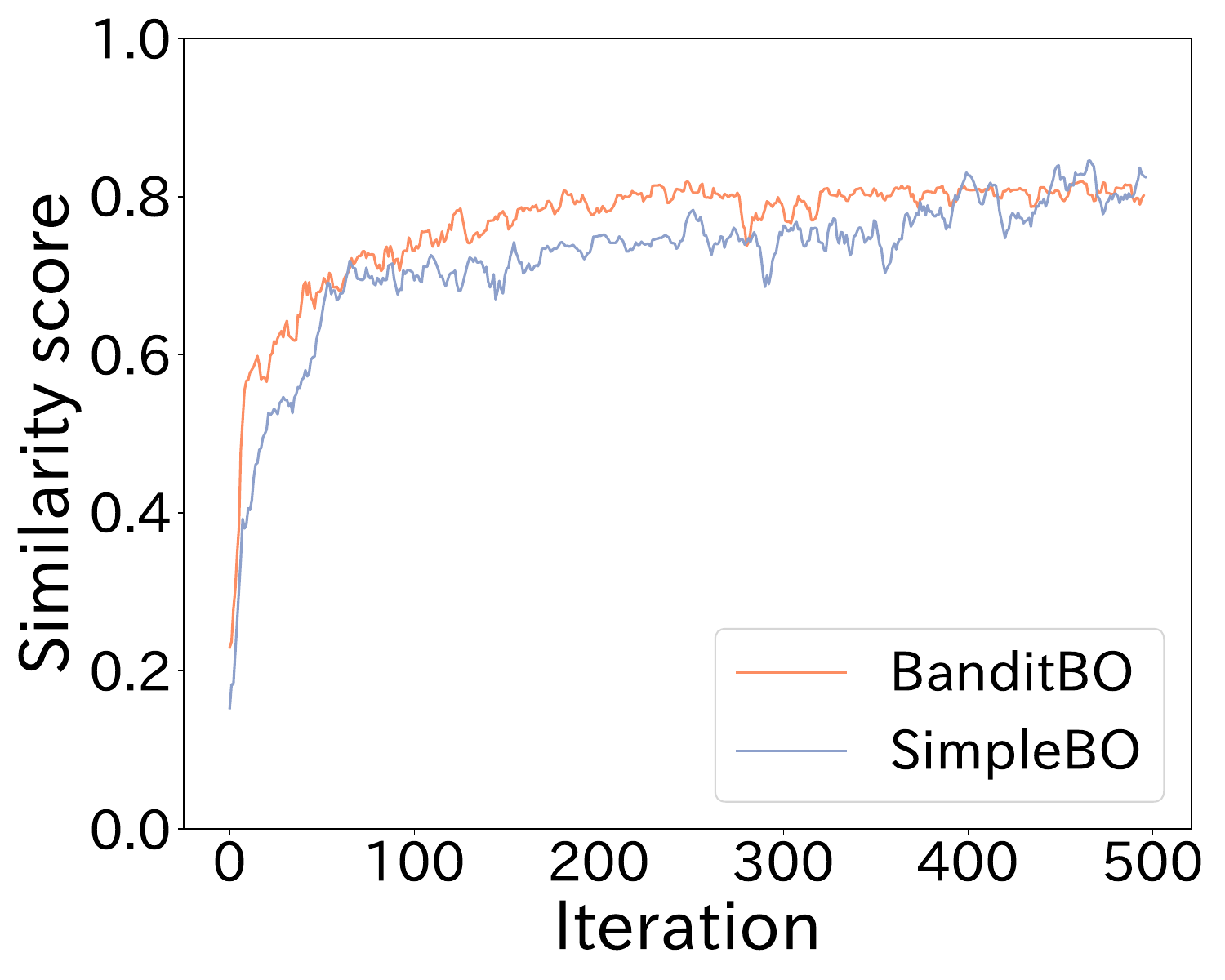}
    \caption{$d^{\prime}=4$}
  \end{subfigure}
  \begin{subfigure}{0.32\linewidth}
    \includegraphics[width=\linewidth]{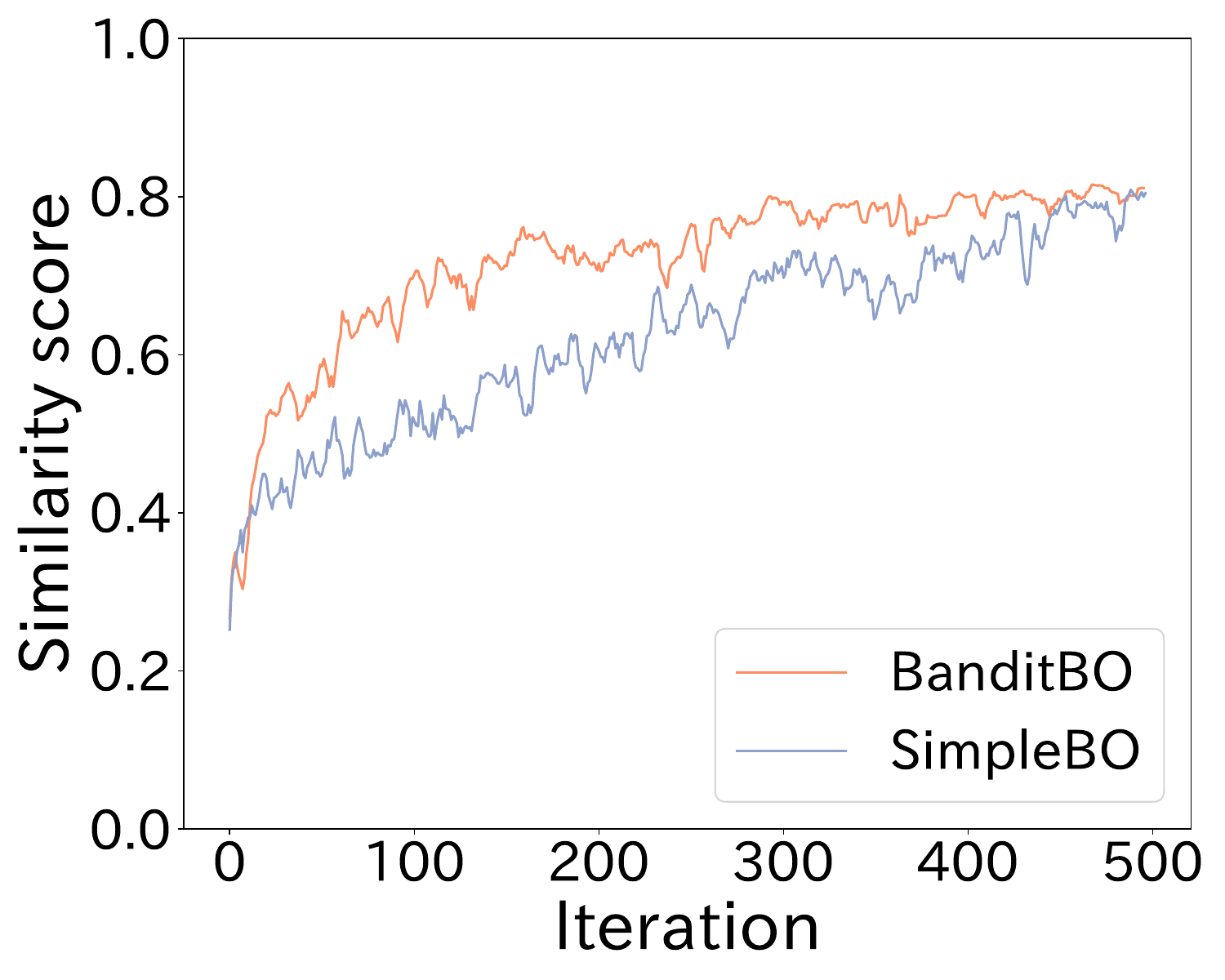}
    \caption{$d^{\prime}=8$}
  \end{subfigure}
  \begin{subfigure}{0.32\linewidth}
    \includegraphics[width=\linewidth]{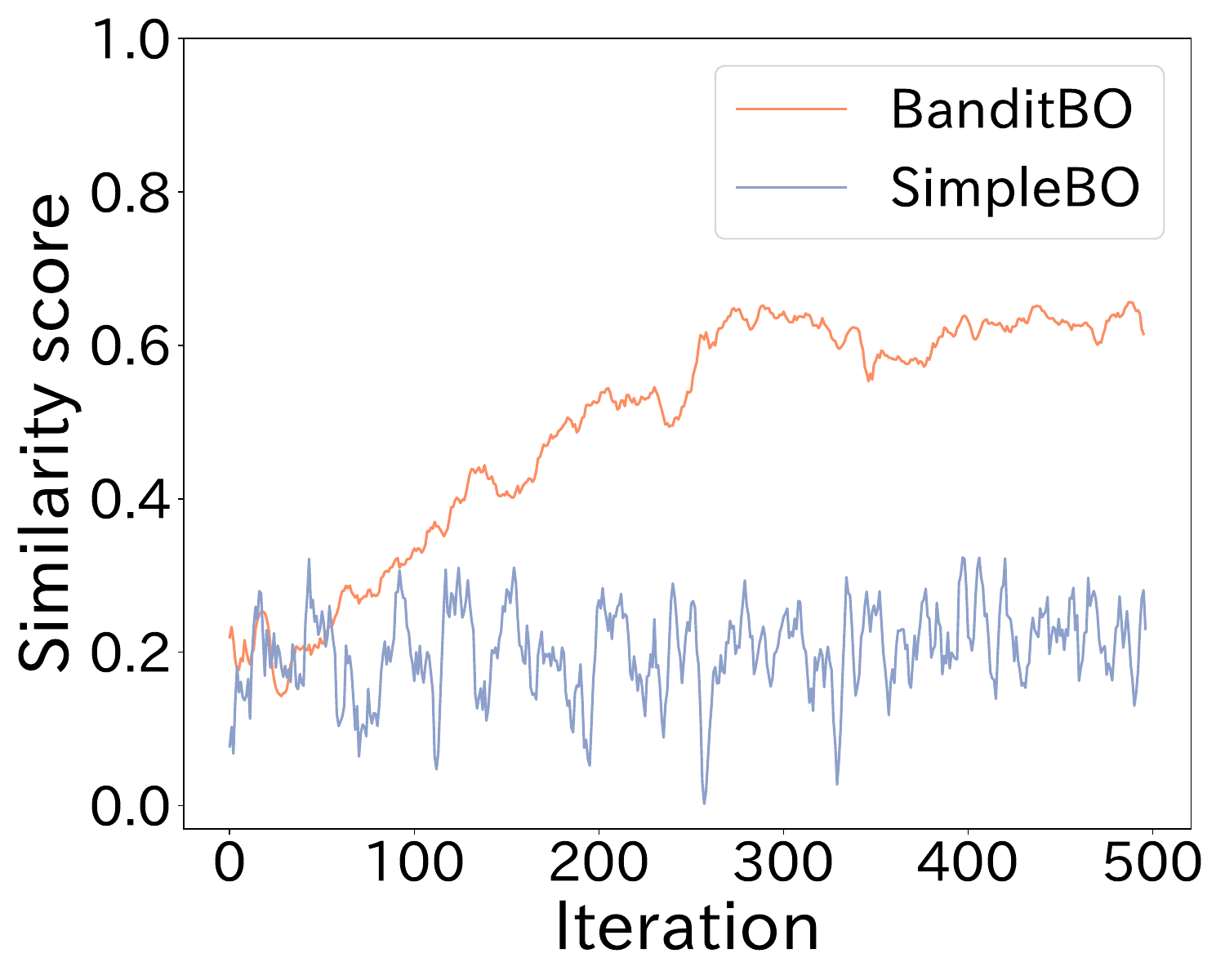}
    \caption{$d^{\prime}=16$}
  \end{subfigure}
  \caption{Trends in the similarity between the generated image and the target image for each $d^{\prime}$(averages of moving averages over $10$ target images).}
  \label{fig: change of facenet score}
\end{figure*}

Figure~\ref{fig: change of facenet score} shows the moving average of the similarity for each $d^{\prime}$. When $d^{\prime}=4$, there was no significant difference between BanditBO and SimpleBO. For $d^{\prime}=8$, BanditBO became closer to the target image more efficiently than SimpleBO compared to the case of $d^{\prime}=4$. Similarly, for $d^{\prime}=16$, BanditBO became even closer to the target image more efficiently than SimpleBO. These results suggest that the proposed method approaches the target image more efficiently than the competing method when the search dimension is high.

%% file: source/body/UserExperiment.tex
\subsection{Experimental setup}
To verify whether the proposed method can efficiently generate preferred images in practical situations than the baseline methods, we conducted user experiments. Participants were asked to generate their preferred avatars for a scenario. We prepared six scenarios: lawyer, teacher, sports instructor, receptionist, library staff, and childcare worker. Each participant was asked to perform all the scenarios and the order of the scenarios was randomly shuffled for each participant. There were three methods: BanditBO, SimpleBO, and Random. The random method samples intermediate latent variables from the search space with a uniform distribution. The dimensionality of the subspace $W^{\prime}$ was chosen from $d^{\prime} \in \{4, 16\}$. There were six conditions (three methods and two dimensions), and each of the six scenarios was randomly assigned to one of them without overlap for each participant. Fourteen participants ($13$ men and $1$ woman, aged $18$ to $27$ years) participated in the study. The participants were told that they could only compare a maximum of fifty images in each scenario. Participants compared images on their PCs, not by swiping on their smartphones.

To let users assume the kind of image they want in advance, they were asked to imagine their preferences before starting each scenario. They were told that they did not necessarily have to generate images as per the preliminary questionnaire. They were also instructed to focus only on the faces for comparison. To evaluate whether the proposed method can efficiently generate preference images compared to the baseline methods, we prepared specific questions designed to gauge if the method efficiently arrived at the preferred images (Q$1$) and to assess the quality of the process of image generation (Q$2$, Q$3$) as well as the impression of the final generated image (Q$4$). The questions are as follows:

\begin{description}
\item [Q1.] Were you able to reach the preferred image efficiently?
\item [Q2.] Did the images presented gradually change to the ones you preferred?
\item [Q3.] Did you find it easy to compare images each time?
\item [Q4.] Were you satisfied with the final image produced?
\end{description}

We employed a seven-point Likert scale for these questionnaires ($1$: Strongly disagree, $7$: Strongly agree), and the participants were asked through free description what they felt in the scenario. After going through all scenarios, participants were also asked to describe in free description what they felt throughout the experiment. Regarding efficiency, if BanditBO can generate preference images efficiently, then the number of image comparisons for BanditBO should be fewer than those of the baseline models. To verify this, we examined the distribution of the number of image comparisons required to generate the preference images. Also, regarding the difficulty of comparing images, if the comparison of images is challenging, it should take more time to select the images. For the time spent on selecting images at each iteration, we visualized the average selection time for each method and dimension.

\subsection{Results}
\subsubsection{Efficiency}

\begin{figure}[tb]
    \centering
    \includegraphics[width=.8\linewidth]{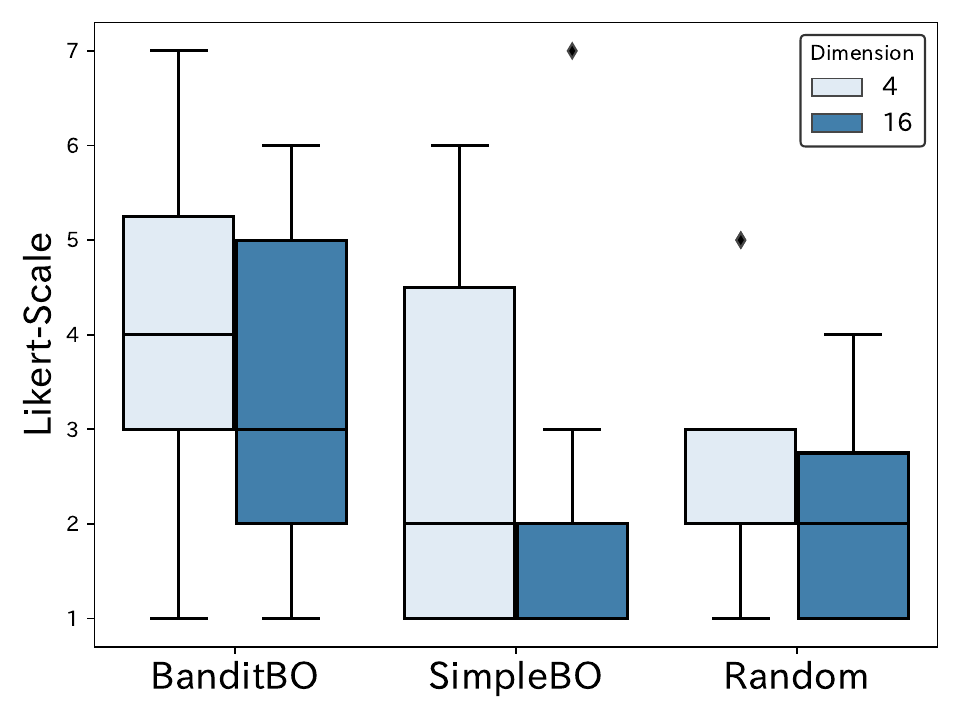}
    \caption{Distribution of response to ``Were you able to reach the preferred image efficiently?''}

    \label{fig:question_efficient}
\end{figure}

\begin{figure}[tb]
    \centering
    \includegraphics[width=.8\linewidth]{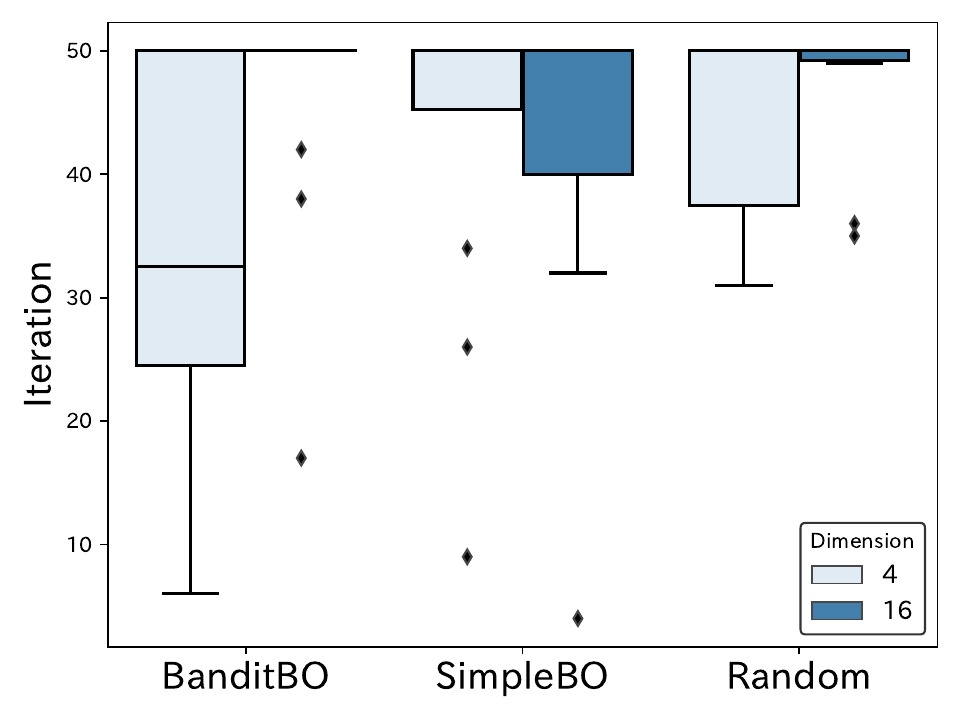}
    \caption{Distribution of the number of image comparisons needed to generate preferred images}

    \label{fig:number_of_comparisons}
\end{figure}

Figure~\ref{fig:question_efficient} shows the results of the question ``Were you able to reach the preferred image efficiently?''  It is observed that BanditBO outperformed the baselines with respect to the median by $2$ points in $d^{\prime}=4$ and $1$ point in $d^{\prime}=8$. Figure~\ref{fig:number_of_comparisons} illustrates the results of the distribution of the number of image comparisons needed to generate preferred images. The performance of BanditBO in the $d^{\prime}=4$ case tends to have the lowest number of comparisons. The free description section in the questionnaire about BanditBO included the following responses:
``It was a good generation, with fine-tuning, comparing significantly different images along the way, and then repeatedly coming back to the original route,'' ``There were various scenarios, such as a scenario where images were appropriately generated and no particular adjustments were made, or another with minor adjustments repeated over a long time, but the best scenario (BanditBO) was a mixture of the two, where I could fine-tune the image and then try one significant shift along the way and move to that scenario if necessary,'' and ``Compared to other scenarios, the changes in the image between iterations were smaller in most cases and seemed to be reached more efficiently.''  From these, it was concluded that the proposed method can generate preferred images more efficiently by appropriately combining small and large changes. For SimpleBO with $d^{\prime}=4$, although there was good feedback at some points, there were free descriptions such as ``The images were difficult to select because most of them hardly changed, and when they did change, the changes were unrelated to the purpose of the image.'' This indicates that SimpleBO dit not change what the user wanted to change.

\subsubsection{User experience}

\begin{figure}[tb]
    \centering
    \includegraphics[width=.8\linewidth]{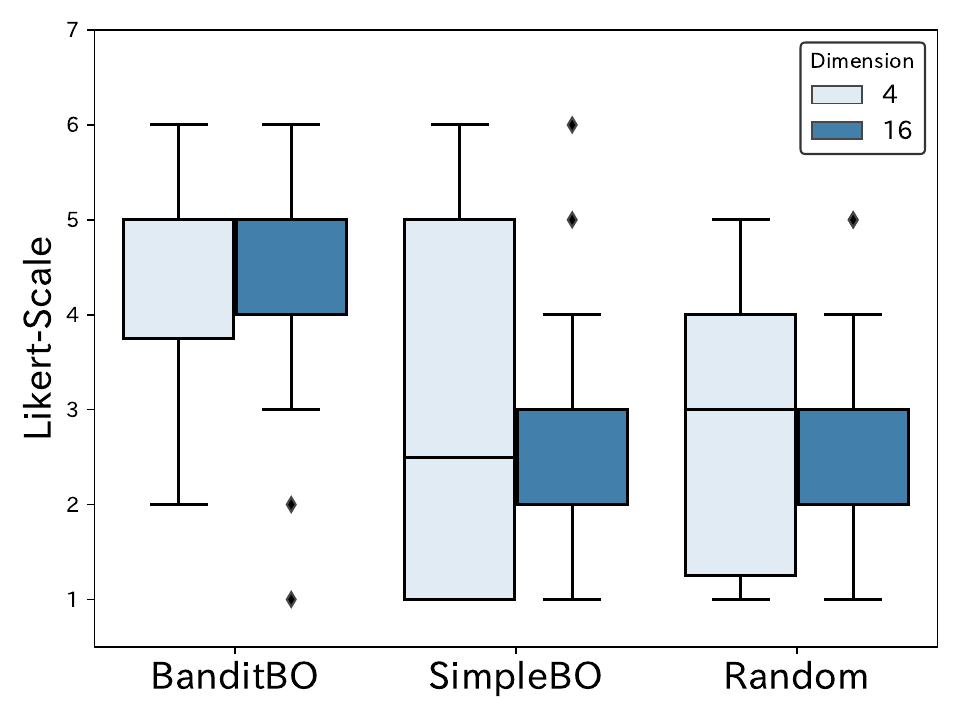}
    \caption{Distribution of response to ``Did the images presented gradually change to the ones you preferred?'' }

    \label{fig:question_gradually_prefer}
\end{figure}

Figure~\ref{fig:question_gradually_prefer} shows the results of the question ``Did the images presented gradually change to the ones you preferred?'' This shows that the respondents evaluated the proposed method more favorably than the baseline methods and that there were many cases where the images changed to the preferred images. Figures~\ref{fig:question_efficient} and~\ref{fig:question_gradually_prefer} show that the performance of SimpleBO in the $d^{\prime}=4$, case is not superior to that of BanditBO. The feedback from users included many comments such as ``It was hard to choose images, as they changed very little most of the time, and when they did, the modifications were usually not related to the goal.'' Consequently, SimpleBO had a negative user experience.

\begin{figure}[tb]
    \centering
    \includegraphics[width=.8\linewidth]{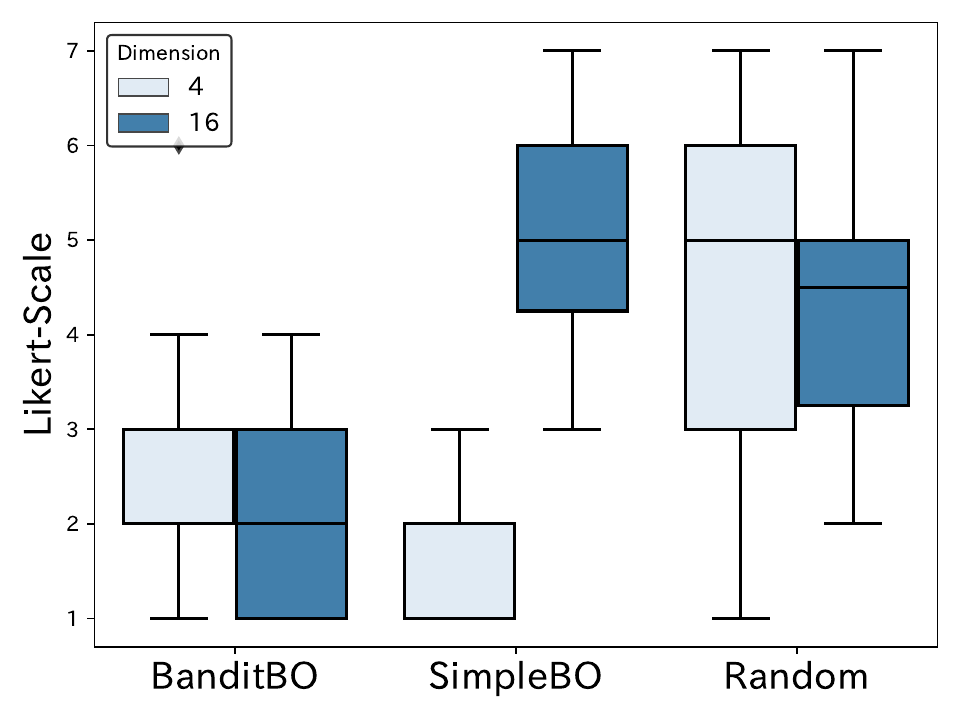}
    \caption{Distribution of response to ``Did you find it easy to compare images each time?'' }

    \label{fig:question_difficulity}
\end{figure}

\begin{figure}[tb]
    \includegraphics[width=.9\linewidth]{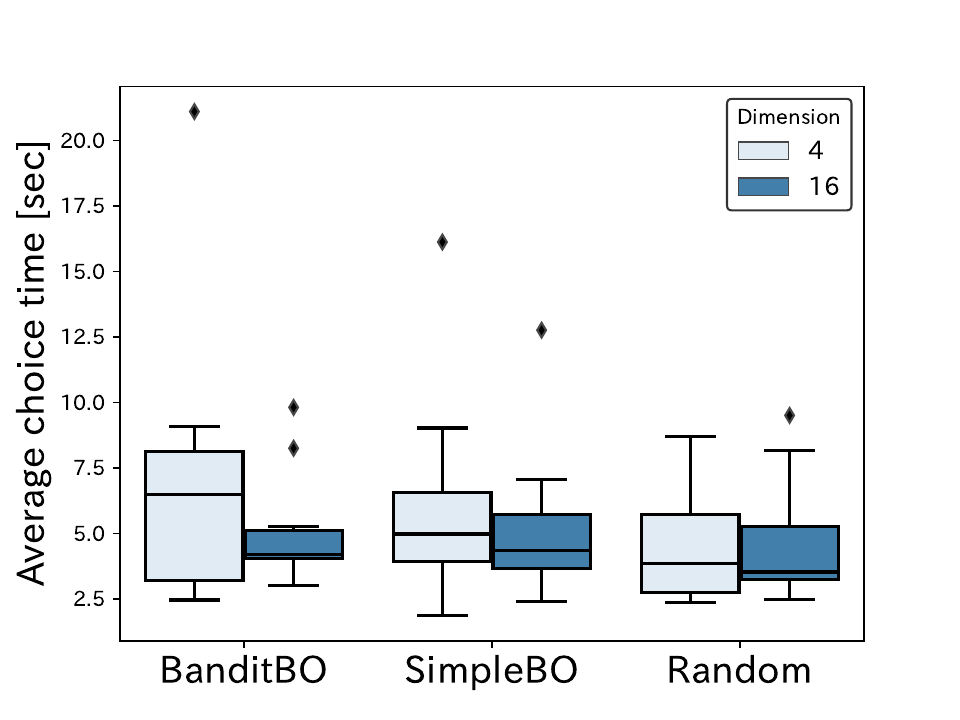}
    \caption{Distribution of average selection time per image} 

    \label{fig:avg_of_choice_time}
\end{figure}

Figure~\ref{fig:question_difficulity} shows the results of the question after each scenario, asking ``Did you find it easy to compare images each time?'' It can be seen that the comparison of images in the $d^{\prime}=4$ cases of BanditBO and SimpleBO was difficult. The responses included many free descriptions, such as ``The same images are generated continuously, making it difficult to select the best image.'' It was found that users did not like comparisons of similar images. Figure~\ref{fig:avg_of_choice_time} shows the distribution of average selection times for each method. BanditBO and SimpleBO have longer average selection times than Random. This may be attributed to the fact that similar images are difficult to compare, and therefore, the selection process takes longer.

\subsubsection{User satisfaction with the final outcome.}
\begin{figure}[tb]
    \centering
    \includegraphics[width=.8\linewidth]{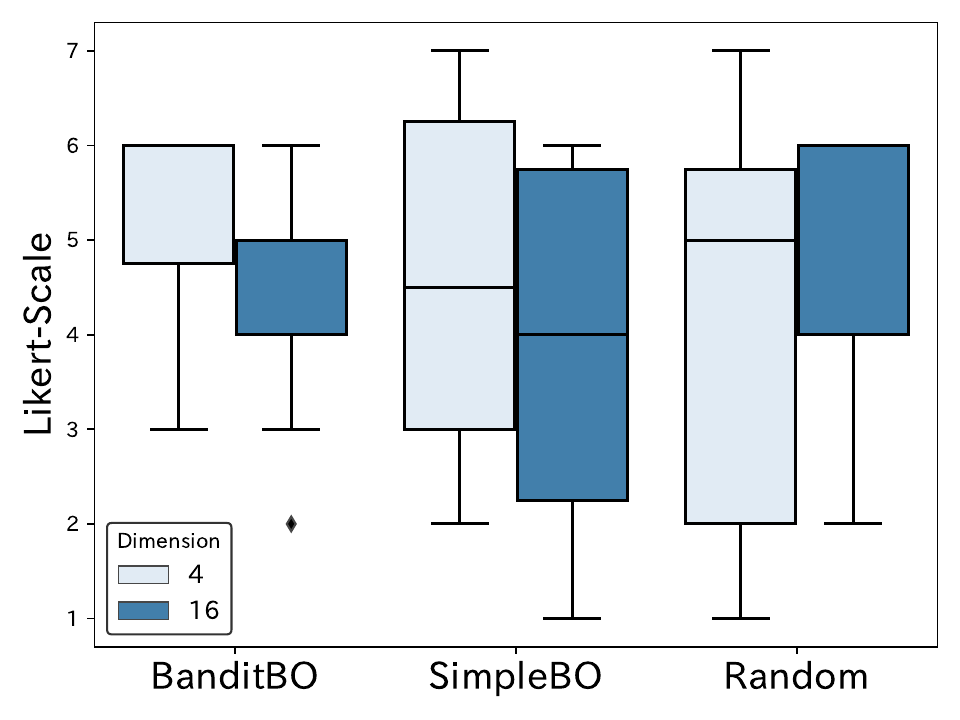}
    \caption{Distribution of response to ``Were you satisfied with the final image produced?'' }
    \label{fig:question_satisfy}
\end{figure}

Figure~\ref{fig:question_satisfy}  shows the results of the question asking ``Were you satisfied with the final image produced?'' after each scenario. This shows that users were more likely to be satisfied with the final image in the proposed method compared to the baseline in the $d^{\prime}=4$ case. The variances in the distribution of results for the SimpleBO and Random questionnaires are greater than that of BanditBO. The results of the questionnaire for the Random method showed both a positive comment that the reason for the system stoppage was ``because the avatar output was close to the image that was answered in the preliminary questionnaire,'' and a negative comment in the free description that ``the image did not change as expected.'' The free description comments for SimpleBO also revealed both positive comments, such as, ``There were some iterations that deviated from my expectations, but in the end I was able to create images that I was satisfied with.'' and negative comments such as, ``There were almost no changes in the images, and it was difficult to compare the results.'' From these results, it is clear that SimpleBO and Random differ in the degree of satisfaction with the final image depending on users.

In summary, BanditBO generates preferred images efficiently, which gradually become closer to the preferred images, and the quality of the final generated image is also good. However, the drawback is that comparing images in each iteration is difficult and time-consuming.

%% file: source/body/Discussion.tex
\subsection{Changes in preferences due to image comparison}
The questionnaire responses included free descriptions such as 
``Gender of the presented image often changed during image generation depending on the occupation of the desired avatar. Also, images that were clearly different from what the user imagined were often generated during the process, but displaying an image that was completely different from the previous ones, sparked a sort of inspiration in changing my preferences of the avatar'' and ``I preferred to create male avatars, but several times women appeared along the way. The appearance of women changed my preferences a little, but on repeating the process of creation, my preference changed to men.'' This suggests that human preferences change during image comparison. The proposed method responds to such changes in user preferences. The multi-armed bandit algorithm dynamically evaluates the contribution of each dimension to the current user preferences, allowing the generation of new images that correspond to changes in user preferences. 

A future approach for efficient response to temporal changes in preferences is to weight rewards by time using a multi-armed bandit algorithm: giving smaller weights to past rewards and larger weights to newer rewards. This method would make the proposed system more sensitive to recent preferences and allow an efficient response to subtle changes in user preferences.

\subsubsection{Promoting broader scope of user exploration in the proposed method}
Previous research associates creative ideas with originality and usefulness~\cite{TheCambridgeHandbookofCreativity}. Previous studies have proposed that in Bayesian optimization, exploration and exploitation are similar to the concepts of originality and usefulness, respectively, possibly bridging the connection with creativity~\cite{BO_as_assistant}. Free descriptions about BanditBO included, ``Two avatars that were quite similar were often presented. Sometimes they would change them a lot, but then they would return to the same picture and repeat fine-tuning,'' and ``It was doing a good job of fine-tuning, repeatedly coming back to the original after comparing it to something significantly different along the way.'' These observations indicate that Bayesian optimization can facilitate creativity. Furthermore, the final questionnaire included free descriptions such as, ``There were various scenarios, such as a scenario where images were appropriately generated and no particular adjustments were made, or another with minor adjustments repeated over a long time, but the best scenario (BanditBO) was a mixture of the two, where I could fine-tune the image and then try one significant shift along the way and move to that scenario if necessary.'' This suggests that the balance between exploration and exploitation of the proposed method is similar to that between originality and usefulness in creative idea generation. In particular, the BanditBO approach enables the retention of originality while pursuing usefulness.

\begin{figure}[tb]
 \centering
  \begin{subfigure}{\linewidth}
    \centering
    \includegraphics[width=0.8\linewidth]{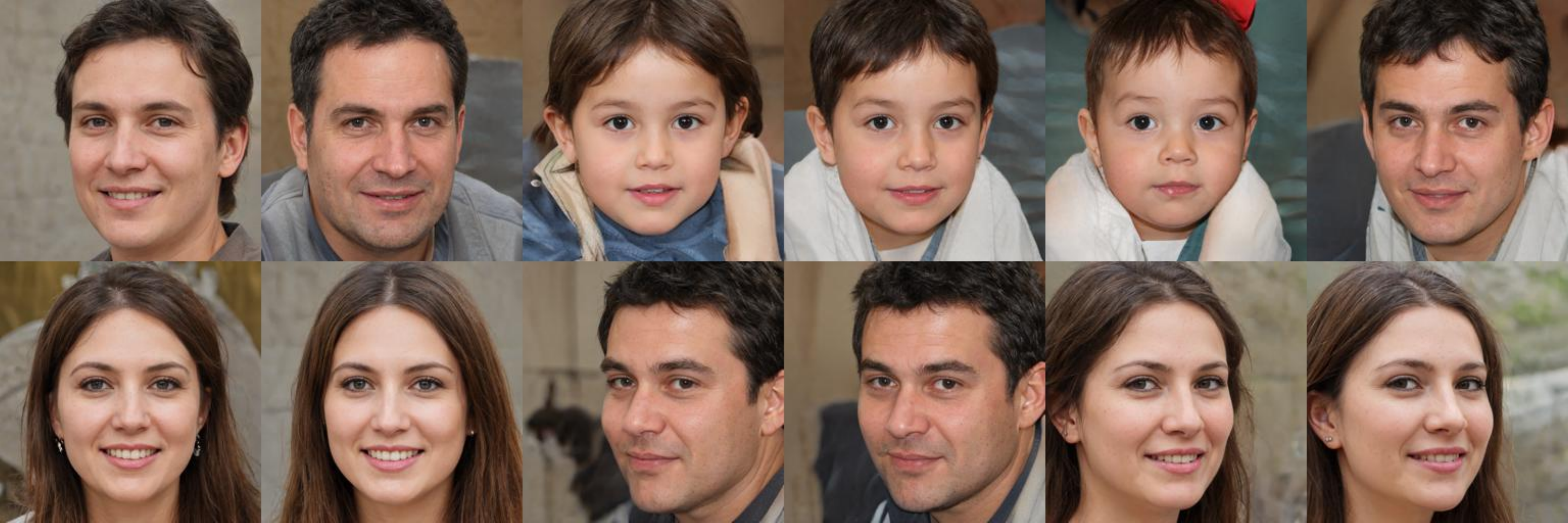}
    \caption{$d^{\prime}=4$}
  \end{subfigure}
  \vfill
  \begin{subfigure}{\linewidth}
    \centering
    \includegraphics[width=0.8\linewidth]{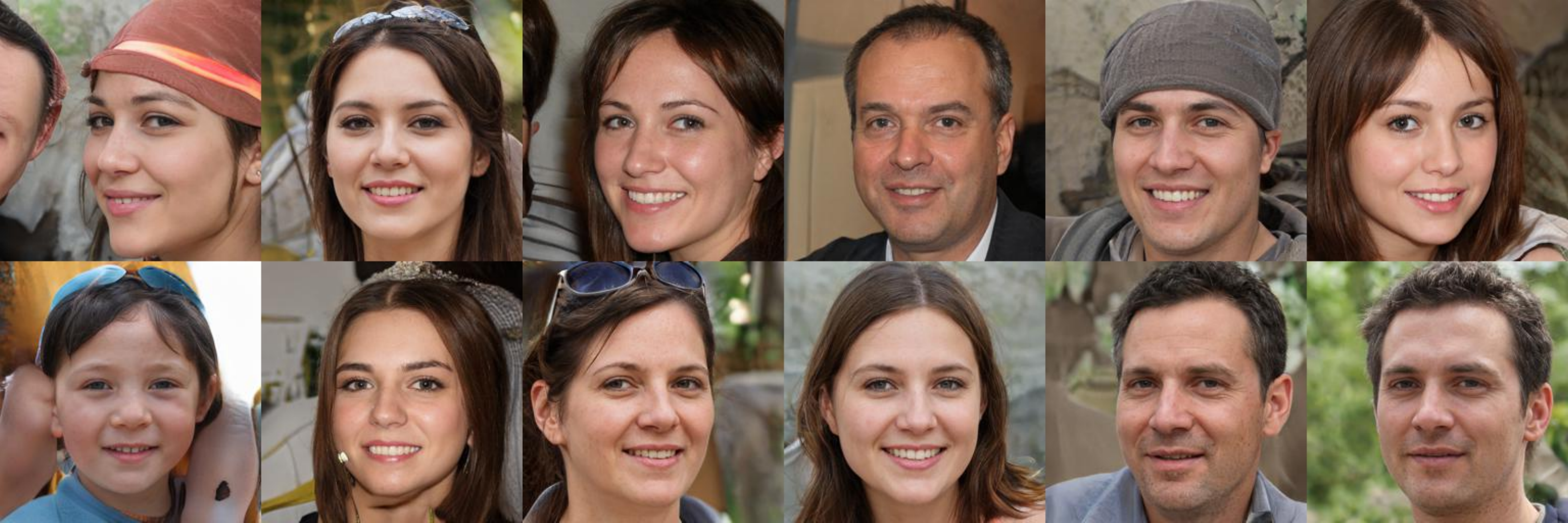}
    \caption{$d^{\prime}=8$}
  \end{subfigure}
  \vfill
  \begin{subfigure}{\linewidth}
    \centering
    \includegraphics[width=0.8\linewidth]{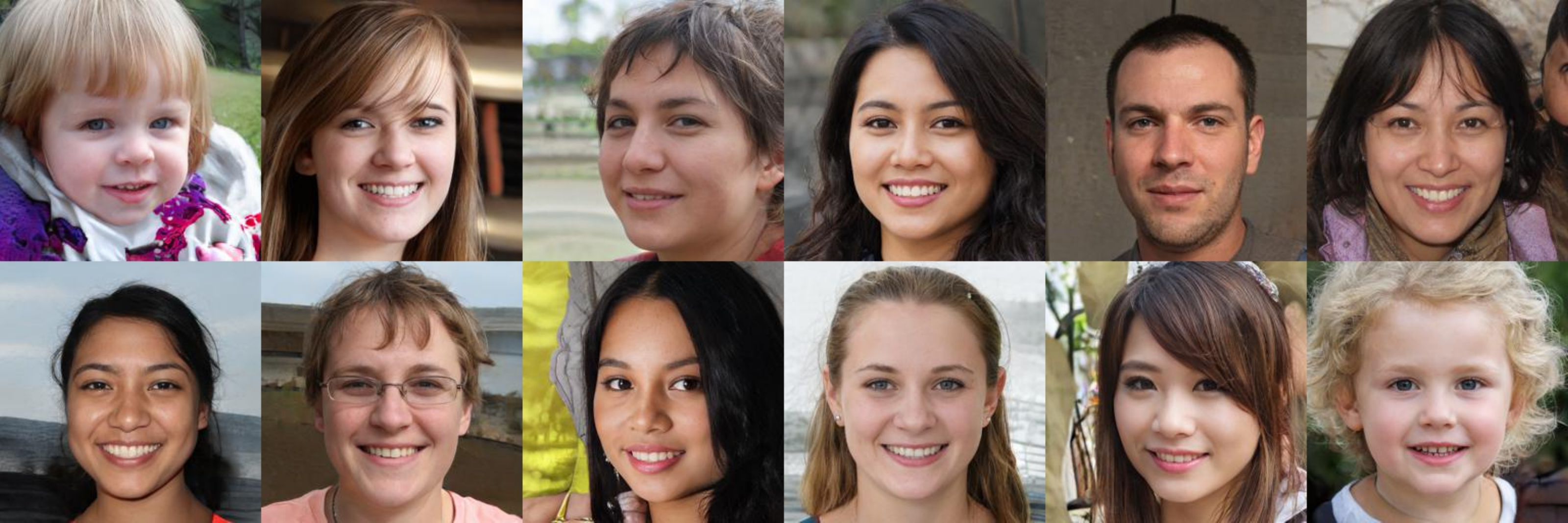}
    \caption{$d^{\prime}=16$}
  \end{subfigure}
  \caption{Verification of the difference in expressivity depending on the dimensionality $d^{\prime}$ of the subspace}

  \label{fig: sample image}
\end{figure}

\subsection{Relationship between diversity of generated images and user satisfaction}
The general feedback from the participants included the following text: ``Scenarios with poor efficiency had images generated that were almost unrelated to each other in sequence, and I felt it did not respond to my comparisons. For scenarios where the efficiency was good, images began to be changed little by little from a small number of images, and I felt that the selections reflected my choices, but the output after that was almost determined by the first selected image, and I felt that it settled into the locally optimal image.'' These comments indicate that the images generated by the proposed method lack diversity compared to those generated by Random. We also believe that the diversity of images differ based on the size of subspace, rather than differences between methods. We assessed the size of the subspace and the expressive power of the GAN by generating multiple images from random samples $w^{\prime}$ for each $d^{\prime}$ to confirm the variety of the generated images in different dimensions $d^{\prime}$.

The results are shown in Figure~\ref{fig: sample image}. A wide variety of images were generated under $d^{\prime}=8, 16$. However, under $d^{\prime}=4$, similar images were generated but were not diverse. Reducing the latent space of StyleGAN using PCA resulted in a qualitative confirmation of the diminished expressiveness in the image generation of StyleGAN. In addition, the free descriptions regarding Random under $d^{\prime}=4$, included several responses like, ``Images seen before often reappeared.'' These results indicate that the latent space with $d^{\prime}=4$ lacked diversity in the generated images compared to the space with $d^{\prime}=16$.

\begin{figure*}[tb]
  \centering
  % First row
  \begin{subfigure}{0.49\linewidth}
    \centering
    \includegraphics[width=.8\linewidth]{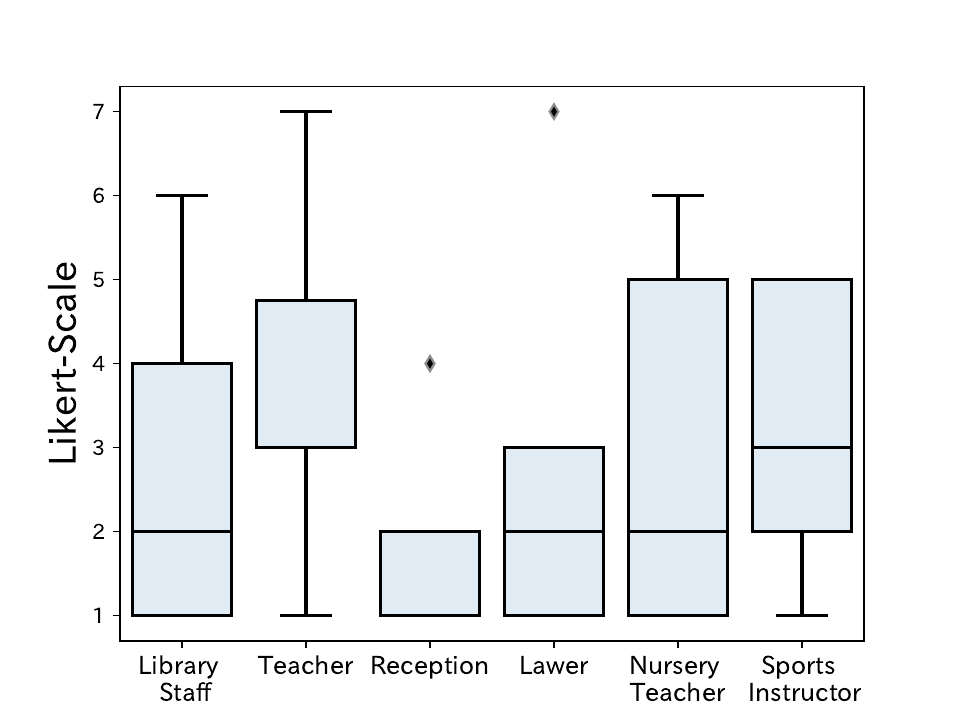}
    \caption{``Were you able to reach the preferred image efficiently?''}
  \end{subfigure}
  \hfill % Space between the first and the second image
  \begin{subfigure}{0.49\linewidth}
    \centering
    \includegraphics[width=.8\linewidth]{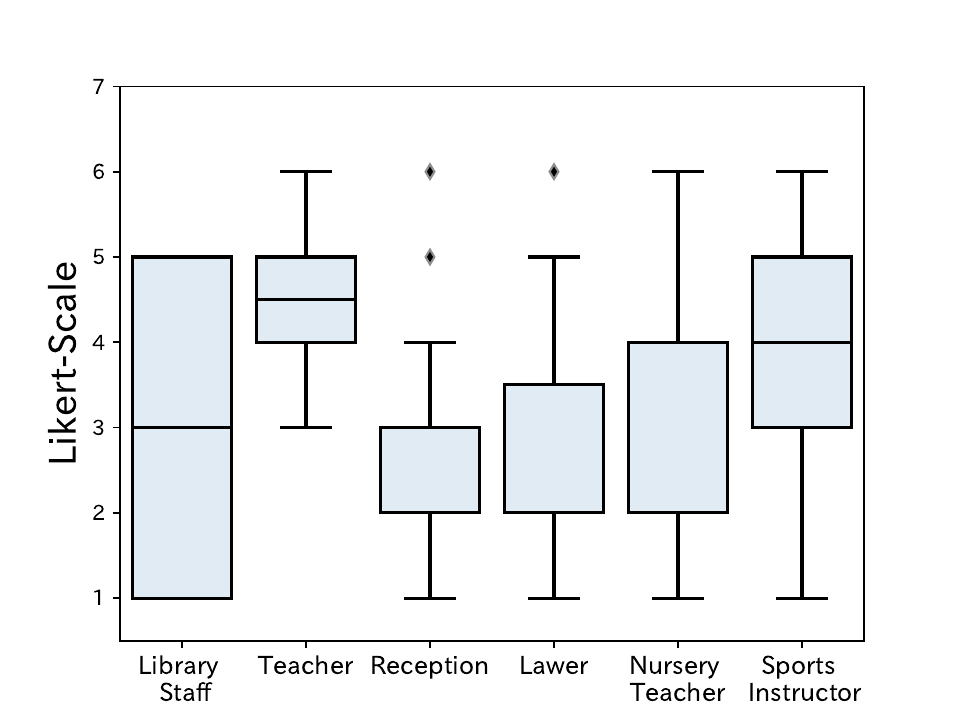}
    \caption{``Did the images presented gradually change to the ones you preferred?''}
  \end{subfigure}

  % Second row
  \begin{subfigure}{0.49\linewidth}
    \centering
    \includegraphics[width=.8\linewidth]{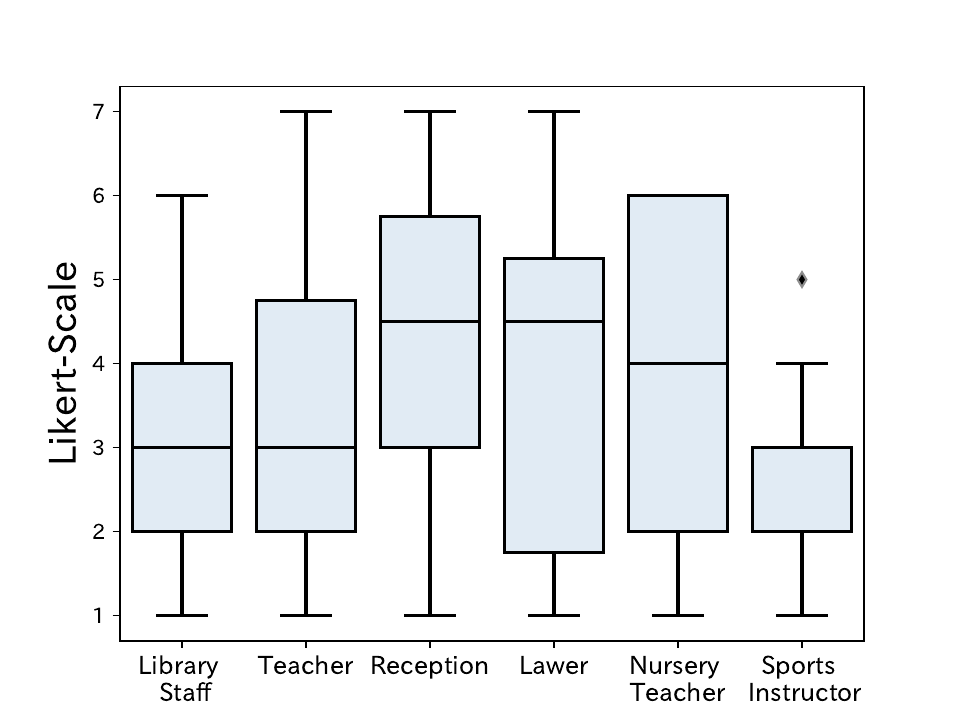}
    \caption{``Did you find it easy to compare images each time?''}
  \end{subfigure}
  \hfill % Space between the third and the fourth image
  \begin{subfigure}{0.49\linewidth}
    \centering
    \includegraphics[width=.8\linewidth]{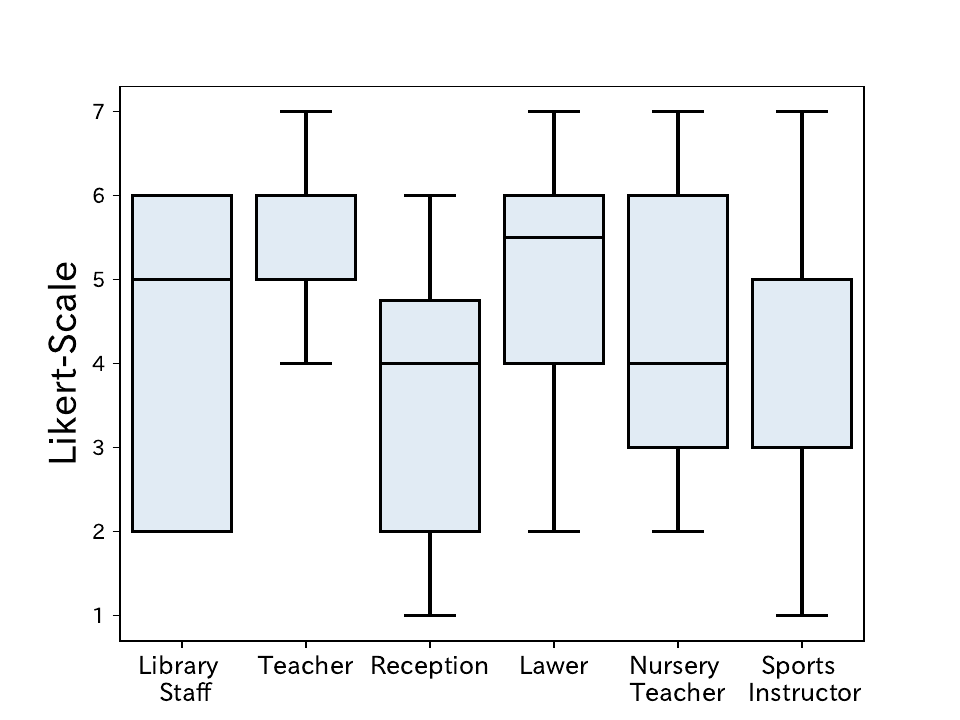}
    \caption{``Were you satisfied with the final image produced''}
  \end{subfigure}

  \caption{Distribution of response to each question in each scenario}

  \label{fig: scenario fig}
\end{figure*}

\begin{figure}[tb]
    \includegraphics[width=.8\linewidth]{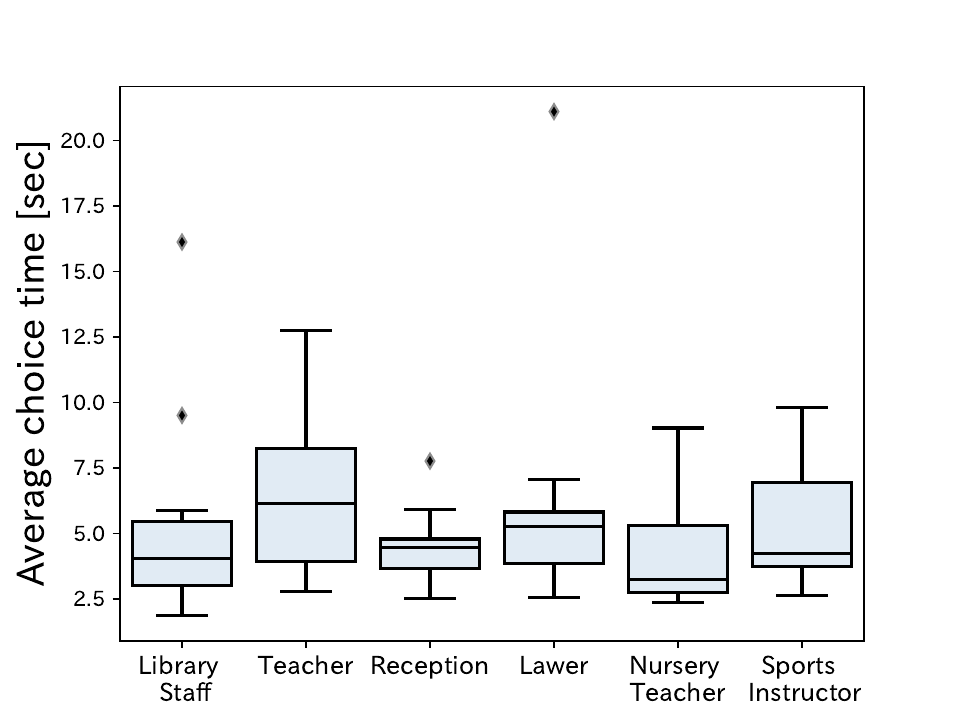}
    \caption{Distribution of average selection time per image in each scenario}
    \label{fig:scenario_avg_choice_time}
\end{figure}

However, responses for both dimensions included comments such as, ``I was able to generate an avatar that was close to the image I answered in the preliminary questionnaire,'' and ``I was able to generate an image that I was satisfied with, even though it was different from my initial image.” This suggests that although differences in dimensionality affect the diversity and expressiveness of the generated images, they may not directly affect the ability to generate images that satisfy the user.

\subsection{Impact of scenario settings}
Figure~\ref{fig: scenario fig} shows the results of the post-scenario questionnaires for each scenario. It shows that (a) users achieved the preferred image of the teacher more efficiently than that of library staff; (b) in the teacher scenario, more images gradually shifted toward preferences, compared to the lawyer scenario; (c) users found it easier to compare images in each iteration for the childcare worker scenario than for the sports instructor and teacher scenarios; and (d) there was higher satisfaction with the final generated images of teachers than with those of sports instructors. These results indicate that the scenarios presented may have influenced the questionnaire. When examining the preliminary questionnaire by scenario, it was noted that for the preliminary questionnaire of lawyers, over $75\%$ 
of the participants used the keyword ``$40$s,'' while for teachers, a wide range of ages, from ``$20$s'' to ``$60$s,'' were mentioned. Furthermore, in the preliminary survey for receptionists and childcare workers, more than $92\%$ of the respondents indicated ``female'' while in those for teachers and sports instructors both ``male'' and ``female'' were included. We gain valuable insight through these observations. First, because the preliminary questionnaire for the teacher scenario included responses from a diverse range of age groups and genders, the diversity and flexibility of images of teachers may have contributed to a more efficient generation of images and user satisfaction. Conversely, uniform keywords in the lawyer, childcare worker, and receptionist scenarios, coupled with instructions to create images according to the scenario, likely influenced the results owing to strong user preconceptions. Figure~\ref{fig:scenario_avg_choice_time} shows the average selection time of the participants per scenario. The average selection time for the childcare worker scenario was shorter than that for teachers. The stereotypes of childcare workers may have made it easier to compare images than in the teacher scenario, where users had various preconceptions. Exploring how these factors impact image accuracy and user satisfaction warrants detailed analysis in future studies.

%% file: source/body/DesignImplication.tex
\subsection{User-oriented systems}
\label{subsec: bo as assistant}
This section discusses several important aspects related to system usability improvements for generating user-preferred images. The questionnaire responses included free descriptions of the content the users wanted adjusted in the system, such as ``I would have liked to have the option of not selecting either of the choices, as being forced to select one of the two images (even if I did not want either) led me in a completely different path,'' ``I think it might be more efficient to skip images that differ significantly from the preferred ones when presented as a two-choice option,'' and ``I wanted to change the glasses aspect, but aspects other than glasses kept changing, and it was difficult to obtain the image I wanted.'' These user responses indicated a preference for autonomy in the system. A previous study proposed a concept where human designers have complete control through Bayesian optimization~\cite{BO_as_assistant}. Experiments have not yet been conducted on this concept; therefore, it is unclear whether users require user-oriented systems. In contrast, our experiment results reveal the need for a user-oriented design. Thus, we propose conducting a demonstration experiment that incorporates user-oriented elements into the system. This study assumed smartphone operation, which imposes a limitation on multiple slider adjustments due to low operability. We need to consider this constraint when examining user-oriented design in system design.

\subsection{Examining comparison images}
The questionnaire responses included several free descriptions such as, ``When the image that I did not choose became the next target of comparison, it felt like the comparison was moving away from my preferred image, which was irritating,'' and ``I thought that having the image we selected in the previous comparison carried over to the next comparison would be a better and easier means of comparison.'' These results reveal that users wanted the system to compare the image selected in a previous comparison with the one generated. However, this study assumed smartphone operation, and given the limitation of the small screen, a user interface with one image per screen was adopted, similar to applications such as Tinder~\footref{tinder}. In this case, we believe it is natural to compare the image with the previously generated image. We need to consider what the optimal user interface for pairwise comparisons on a smartphone would be.

\subsection{Visualizing image changes}
In the questionnaire responses, we received the free description: ``It is often difficult to understand how the two images differ in terms of fine-tuning. If you are offering this on a feature basis, it would be helpful if you provide a small description of how the two differ.'' This suggests the need to visualize the parts of the newly generated image that have been changed compared with the previous image. One way to address this need is to calculate the difference between the two at the pixel or feature levels and display the difference as a heatmap or mask. This could show the changed parts at a glance. Applying existing feature extraction algorithms to the generated images and comparing the results would also be useful. Such a feature-based approach should help capture differences at a higher semantic level.

\subsection{Showing the balance between exploration and exploitation}
The questionnaire responses had the free description, ``I feel that if I were informed whether my choices are narrowing the image roughly, or whether I was selecting detailed features when selecting images, it would reduce my anxiety about the selection.'' To mitigate user anxiety, we found it useful to display the current balance between the exploration and exploitation of Bayesian optimization to users. As described in Section~\ref{subsec: bo as assistant}, allowing users to adjust the balance between the exploration and exploitation of Bayesian optimization is a prospect for the future. However, designers must still devise ways to facilitate this adjustment and problems of interface complexity, and helping users acquire a sense of exploration and exploitation still needs to be addressed.

%% file: source/body/Conclusion.tex
We demonstrated that our proposed method can generate preference images more efficiently than competing methods with only pairwise comparison results. By applying PCA to the latent space of StyleGANs to reduce the dimensions and focusing on the dimensions of interest to the user, the generation of preference images was found to be more efficient than that of baselines. This paper also proposed a new approach for generating preference images using smartphone operation. 

Future work includes developing efficient search methods to generate preferred images with a lower number of iterations, even with a larger number of search dimensions. Specifically, by using additional information, such as the elapsed time from image presentation to swiping, we aim to reduce the number of iterations required to generate the preferred image. Future work also includes making the system more user friendly.